\documentclass[sigconf]{acmart}
\AtBeginDocument{%
  \providecommand\BibTeX{{%
    \normalfont B\kern-0.5em{\scshape i\kern-0.25em b}\kern-0.8em\TeX}}}

\setcopyright{acmcopyright}
\copyrightyear{2018}
\acmYear{2018}
\acmDOI{XXXXXXX.XXXXXXX}

\acmConference[CSCW 2022]{CSCW}{November,
  2022}{Situating Network Infrastructure with People, Practices, and Beyond}
\acmPrice{15.00}
\acmISBN{978-1-4503-XXXX-X/18/06}

\begin{document}

\title{Lessons from Digital India for the Right to Internet Access}

\author{Kaustubh D. Dhole}
\email{kaustubh.dhole@emory.edu}
\affiliation{%
    \institution{Department of Computer Science}
  \institution{Emory University}
  \streetaddress{P.O. Box 1212}
  \city{Atlanta}
  \state{Georgia}
  \country{USA}
  \postcode{43017-6221}
}

\begin{abstract}
With only 65\% of Indian houses having access to the Internet, digital India faces a significant Internet divide across gender and city types. Rendering essential services inaccessible to almost a third of the population necessitates not only provisioning a fundamental right to Internet access but taking specific constructive steps to assure its simple, affordable and safe accessibility. Establishing such a right would also pave way for other ancillary rights required for data privacy, protection from Internet’s possible harms and the requirement to be treated fairly. We first discuss two arguments on the universal right to Internet access; from Merten Reglitz, a senior lecturer on Global Ethics and from Vincent Cerf, one of the founding creators of the Internet who has had a profound influence on the field. We specifically argue why Internet access should be treated as a fundamental right. We discuss the learnings from India, contextualizing them with the global debate and argue for establishing Internet access as a fundamental right in India and elsewhere in the form of government legislation to eliminate Internet divide.
\end{abstract}





\maketitle

\section{Introduction}
\begin{figure}
    \centering
    \includegraphics[width=\columnwidth]{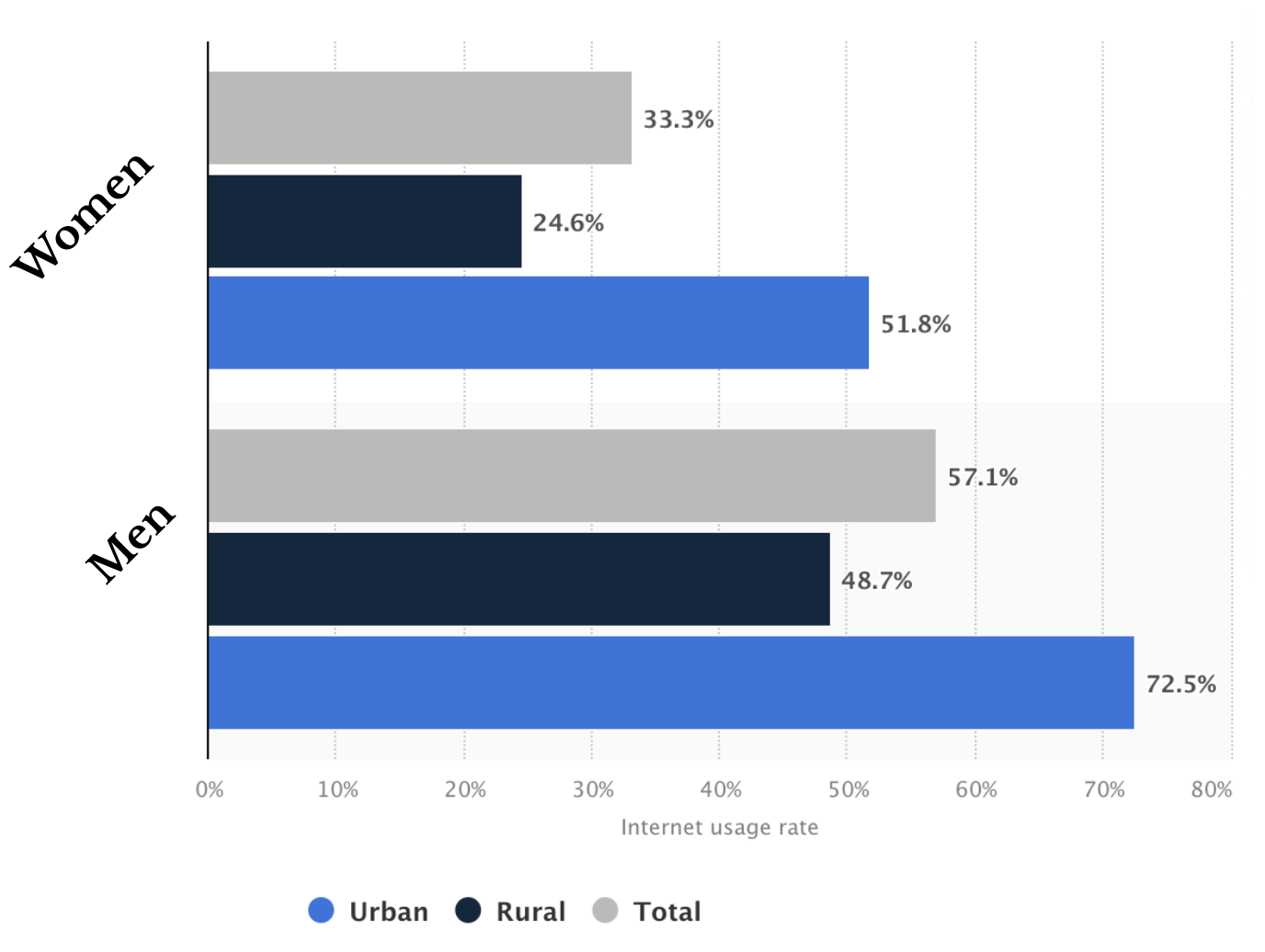}
    \caption{Gender \& City wise Internet usage rate in India between 2019 and 2021~\cite{statista-gender}}
    \label{fig:gender}
\end{figure}

Many of India's digital plans including vaccine distribution, online education, and socio-economic welfare premised on the assumption of widespread and uniform Internet connectivity with high average bandwidth. Online payments were meant to be simple to switch to after India's demonetisation. However, with only 65\% of Indian houses having access to the Internet, digital India faced a significant Internet divide along various lines including gender and city types (Figure~\ref{fig:gender}). This renders essential services inaccessible to almost a third of the population and motivates both establishing a right to Internet access and taking specific constructive steps to assure its easy accessibility. Establishing such a right would also pave the way for other auxiliary rights required for data privacy, protection from Internet's possible harms, and the requirement to be treated fairly. Due to our society's shift towards digitisation and the accompanying dependence on the Internet, it is crucial to not just acknowledge the significance of the Internet but also to campaign for universal access to it in India as well as across the world.

Beyond India, Internet divides manifest in most parts of the world, especially in regions which are highly populated but partially digitised. However, India's case is interesting and globally relevant for a variety of reasons. First, India is a functioning democracy with independent judicial, executive and legislative organs, and hence many of India's domestic implications can be generalised to many other democratic nations. Second, India has both, a large user base with Internet access as well as one without. This contrast provides diverse use cases to study such as large populations missing out on government schemes to large populations accessing affordable Internet. Finally, as we shall see, numerous Indian courts have already passed judgements pertaining to the Right to access the Internet, thereby advancing the global debate of the right further.

Our contributions in this article are two-fold. 1) We first discuss two arguments on the universal right to Internet access from Merten Reglitz, a senior lecturer on Global Ethics and Vincent Cerf, one of the founding creators of the Internet who has had a profound influence on the field. We specifically argue why Internet access should be treated as a fundamental right. 2) We then discuss the learnings from India, contextualising them with the global debate and argue for establishing Internet access as a fundamental right in India and elsewhere in the form of government legislation.

\section{The Larger Debate}
\label{sec}
Whether Internet access should be considered a human right or not has been debated quite extensively in the last couple of years. Apart from large national and international organizations like the United Nations~\cite{unhuman}, Merten Reglitz~\cite{reglitz2020human}, Tim Berners Lee\cite{berners2018Internet} and Vincent Cerf~\cite{cerf2017notaright} were among the forefront of this debate each providing unique perspectives. Let us consider the thoughts of Reglitz and Cerf first which positioned themselves on the opposite sides of the debate.

\subsection{Reglitz's arguments for the Right to Internet Access} \citet{reglitz2020human} argues that Internet access should be considered a universal and moral human right and, everyone should have unsupervised and uncensored access to it. He urges for it to be made available publicly free of charge to those who cannot afford it. Internet access, according to Regtlitz, is more than mere luxury or special privilege. It is necessary to lead a minimally decent life and hence it should be considered a universal entitlement. 

Importantly, he reasons the Internet to be pragmatically essential to protect and promote other well-established and uncontroversial basic human rights like free speech and free assembly. Additionally, the Internet is key to advance democracy for example by its potential to hold global players accountable. The level of impact is such that once Internet access is widely available, a lack of access undesirably excludes people from having meaningful political influence. As Reglitz puts it, \textit{``Without such access, many people lack a meaningful way to influence and hold accountable supranational rule-makers and institutions. These individuals simply don't have a say in the making of the rules they must obey and which shape their life chances.''} There have been a plethora of studies which measure the impact of internet access on political participation. For example, \citet{zhuravskaya2020political}'s study found that free Internet (and social media) do enhance accountability by informing the public and helping the coordination of protests in places where the main public concerns are related to corruption, subversion of power, and autocratic control of conventional media.\citet{shandler2020can}'s early findings indicated that the absence of Internet connection considerably reduced task completion for activities related to political expression and political association and conditionally affected task completion for freedom of information related activities. Why wouldn't then, a lack of internet access itself make for a strong argument for Internet access to be deemed a right?

Besides, the setup costs are not extremely burdensome according to Reglitz - accessing politically vital opportunities on the Internet like blogging and news does not require sophisticated networks such as 5G or glass fibre connections. We will see in the next sections how an Indian telecommunications firm provided Internet access at extremely affordable rates. While there are strong concerns of misuses of the internet, the wide applicability can itself serve as a solution to contain many of them.

\subsection{Cerf's arguments against the Right to Internet Access} \citet{cerf2017notaright}, on the contrary, argues that besides having a high bar for something to be considered as a human right, it should also be fundamental enough, like the things we as humans need in order to lead healthy, meaningful lives, like freedom from torture or freedom of conscience. According to Cerf, the Internet does not pass the high bar to qualify as a human right and is not generic or fundamental enough. Cerf argues that, at some point in the past, it was hard to make a living without horses, but we recognized the right to the generic phenomena of ``making a living'' rather than the right to have a horse. As per Cerf, the Internet acts more as a valuable \textit{means} for critical freedoms of speech and access to information but is not an \textit{end} in itself. When we accept that telephone access must be available everywhere by government policy i.e. as a civil right and not necessarily as a human right, which the UN conflates, Cerf asserts that we analogously must consider the same for Internet access too. 

\subsection{Problems with Cerf's arguments} The Internet is vast and multidimensional and far from being an analog of a horse. Cerf’s analogy to compare the Internet with the horse is a bit misleading. The horse is an extremely specific object which pursues precise goals of say, faster transportation (in the context of old times). The closest Internet counterpart today, albeit in hindsight, probably is a specific website and not the whole Internet, which is clearly more multi-dimensional and a not a mere efficiency enhancing technology. Besides hosting multifarious webpages, it enables access to a plethora of even other technologies as well as exponentially faster communication, smooth reach and influence in cultural and socio-political circles. So, it will doubtlessly render the unreached segment of the population handicapped. Cerf’s insisting on the Internet as just another means treats the Internet as a minor increment over other traditional means like a pen or a physical newspaper and ignores the exponential rapidity, enormity and influence of it.

\subsection{From the lens of Moral philosophy}
\textbf{Deontology}: 18th century German philosopher Immanuel Kant, was famous for his conceptions of deontology. He defined the Categorical Imperative as a commandment or moral code that everyone must follow regardless of their desires or extenuating circumstances. Kant's categorical imperative looked at the will or intention motivating an action to judge the action's morality.

\textbf{Rule Utilitarianism}: On the other hand, rule utilitarianism, another line of consequentialist ethical theory, decided that an action was morally right if it could conform to a rule that would lead to the greatest good. Rule utilitarianism judged an action on the basis of its consequences.

So, can we further take these arguments from moral philosophy to make a clear case whether there should or shouldn't be a human right to Internet access? Kant's Categorical Imperatives are principles which are intrinsically valid and should be obeyed in all circumstances. Kant looks at the will motivating the action. It's not clear if we can compartmentalize the intention of providing internet access into exactly good or bad - The Internet could be used for both, say for performing an online charity fundraiser as well as for promoting genocide. Kant would probably not have been in agreement to vouch for a universal right where the intentions seem hazy. 

However, the consequences of providing the right seem to be alluding towards a more decisive picture as compared to the intentions. Reglitz and Cerf both laud the positive impact that Internet has had on people's lives. Besides, not provisioning a right has had a plethora of negative consequences too. Rule utilitarianism, would find strong reasons to equate the provision of the right to Internet access as an increase in overall happiness.

It is no doubt that Internet access has immense transformative potential and to deny any segment of population access would be to deny them a lifestyle needed to survive a minimum life.

\section{The Indian Context}
Now that we have hopefully established that the right to Internet access is crucial and necessary from a theoretical point of view, let us look at some of its arguments as well as implications in a more practical setting, particularly in the Indian context. We will see why India's case furthers the argument in favour of the right to Internet access. Moreover, India's digital endeavours treat Internet access as a civil right. The next subsections will elaborate why such a civil right conception can be insufficient.  

\subsection{Digital Divide}
As of March 2022, according to the Telecom Regulatory Authority of India~\cite{trai}, just close to 60 percent of India has Internet access. But more importantly, almost 100 percent of urban population has access against 37 percent rural. Moreover, Internet access is concentrated in the top 30 cities. With 57.1 percent of the male population and 33.3 percent of the female population, men used the Internet more often than women in India, a whopping gender gap of 22 percent\cite{statista-gender}. The gender gap is larger for people who have internet access. A male Indian is twice as likely as a female Indian to have Internet access. Such massive divides along lines of gender and city types can easily put millions at a disadvantage. 

In January 2022, 10 school students, hailing from remote villages  of Gadchiroli of the state of Maharashtra complained to the state High Court about lack of access of Internet affecting their studies. The court probed the Department of Telecom (DOT) to gather details. After thorough investigation, the judges released a statement, ``The DoT reply brings on record a very disturbing picture about Gadchiroli where not less than 829 villages lack mobile Internet connectivity. If this is the situation prevailing there, one can very well imagine what would happen to the future of Maharashtra’s next generation which would come from this district. No physical classes are being held and children from these 829 villages are also unable to participate in the digital classes''.

Such a case of digital divide would not be uncommon to witness around the world where other rights would be sacrificed due to Internet divides.\cite{hampton2020broadband}'s survey focusing on rural broadband in 15 Michigan school districts, discovered that children who rely solely on a cell phone for access or who do not have access to the Internet at home perform worse on a number of measures, including digital abilities, homework completion, and grade point average. Additionally, they are less likely to plan to get a college or a university degree. No wonder, Reglitz pointed out, that in such a virtual world some forms of freedom of expression and freedom of assembly are now only accessible via the Internet and those without access are unjustifiably disadvantaged when exercising these rights.

\subsection{Internet Lockdowns}
With regards to Internet lockdowns, most governments have always argued to have taken utilitarian stands i.e. with greater long-term good in mind. Indian laws include the provision to shut down the Internet in the interest of public safety or events of emergency. In the last 10 years, India was subjected to a total of 682 lockdowns~\cite{Internet-shutdowns} each implemented in a different region for different time spans and for different reasons. While some of them  exercised in the states of Assam and New Delhi were intended to curb protests against the government's citizenship amendment bill, some of them were preventive in nature in anticipation of violence. Eg, terrorist activity in parts of the state of Jammu \& Kashmir would always be intensified during any major political event involving cases of stone pelting to terrorists attacking civilians. In August 2019, when the government was about to abrogate, a crucial article, viz. Article 370 in the Indian constitution, the government instituted an Internet ban across the state to prevent .

If we were to make a conceptual distinction of lockdowns using traditional ethical methodology, it would be important to first delineate two types of Internet lockdowns - \textit{preventive lockdowns}, exercised in anticipation of possible violence or political unrest and \textit{emergency or reactive lockdowns}, exercised to suppress political unrest or stop communication post witnessing an event. Let us consider the case of the lockdown during the \textit{Abrogation of Article 370}. If we were to treat provisioning of access as a moral act, deontologists might not be happy if such preventive lockdowns are permitted, as it would seem immoral to restrict Internet access. On the contrary, a consequentialist might argue for reducing possible violence and loss of lives and hence, allow authorisation of such a preventive lockdown. In the cases of reactive lockdowns, the consequences are not obscured and it would not be hard for the deontological and consequentialist forms to come to terms. Eg. when the computers within a network have been infected with a data stealing malware, a temporary shutting off of Internet could be justified.

Preventive lockdowns have been a contentious issue in India. The very serious objections of utilitarianism, are raised against the case of the Abrogation of Article 370; Utilitarianism, specifically Rule Utilitarianism disregards freedom and permits extremely unequal distributions of value, excusing the pain of the disadvantaged i.e. the ones whose internet access was revoked by higher overall advantage viz. reduced loss of lives. Nonetheless, when Anuradha Bhasin filed a petition challenging the constitutionality of the lockdown during the abrogation, the Supreme Court heard the case and also took a utilitarian stand, recognizing the need to balance national security but still gave orders to restore Internet services.

\subsection{Arbitrary Revoking of Access}
In the famous \textit{Faheema Shirin vs State of Kerala case of 2019}, Faheema Shirin\cite{shirin}, a resident at a women's hostel challenged new regulations at her hostel which restricted the use of mobile phones and laptops. The High Court of Kerala declared that the restriction was an unreasonable infringement upon the right to access the Internet, the right to privacy, and the right to education. This case was important for a couple of reasons; first, that Internet restrictions need not be exercised only for political benefit but even for arbitrary reasons; and second, how other crucial rights are heavily dependent on the Internet. 

\subsection{Derivability of Other Rights?}
Less sophisticated alternatives to the Internet can be exploited to curb other derived rights like the Right to Free Speech. It is no surprise that a lot of other rights are intertwined with and derived from the right to Internet Access. The connection with other rights can itself be abused for power without a separate explicit provision. Despite the positive verdict of \textit{Faheema Shirin's case}, it would not be hard to think of a situation in which an Internet regulating body, revokes Internet access in order to restrict another derived right but is not liable since less sophisticated alternatives might exist to still fulfill the right for the purpose of bookkeeping. Even with regards to the lockdown during the abrogation, when the Supreme Court ordered lifting of the Internet shutdown to not restrict the Right to Free Speech, the government quickly restored Internet services, albeit, the speed was throttled to 2G. 

The throttling during the Abrogation of Article 370 too would be viewed differently from deontological as well as utilitarian points of view. At the juncture when the Right to Internet Access assumes a separate right, deontology would look at it as a conflict between the citizen's safety i.e. their Right to Life and the Right to Internet, despite having Internet at 2G speeds. The utilitarian point of view would rather permit such throttling by prioritising the Right to Life over the Right to Internet. Note that this does not invalidate for Internet Access to deserve a special right. It in fact, acknowledges that Internet regulations are necessary, especially when in conflict with other rights. 

It would be hard to disagree with Cerf, when he pointed out that we rather optimize rights derived from the Internet like critical freedoms of speech and access to information. But there is a caveat here - It can hardly happen when we treat the Internet as ``substitutable'' with other alternatives or simply as ``just another means'' in Cerf's terminology. Undeniably, Internet access is vast, unique and non-substitutable, but still not worthy of deregulation.

\subsection{Zero Liability of Private Players}
India's Internet grew exponentially due to government efforts coupled with private players. While private investment is what boosted India's Internet connectivity, it still does not exempt governments from ensuring universal access. Private entrepreneurs and businesses can be unreliable as private businesses would generally only invest when it makes sense economically. A substantive government obligation would in fact put appropriate pressure for governments to deliver without expecting monetary benefits.

\section{Concerns Post the Right}
While we establish India's challenges and present how those justify a separate right to Internet Access, it is important that as a moral right we also justify the relevant duties that its realisation would impose on others. A few concerns have been popularly voiced against the right to Internet Access or at least in general against Internet's current technological form and it is utmost critical to try to address them. 
\begin{itemize}

\item First, is it in our interest to invest and maintain large amounts of physical resources like cables, broadband connections, etc. to provide Internet access to the remaining half of the world today?

\item Second, is it also worth bearing the cost of misinformation harms, concerns of data privacy and other harms that come with the Internet? 

\item Third, whether the realisation of such a right would involve compromising other rights or even undermining other efforts towards improving living standards like electrification? Eg, Shouldn't we first address the global electrification divide when nearly a billion people\cite{energyaccess} in the world are without electrical power?

\end{itemize}

\subsection{Setup Costs} Cerf stated that the positive act of providing Internet access would be too onerous while Reglitz on the other hand argued that the setup costs to have universal Internet access are not so burdensome. The story of one telecommunications private firm in India, Reliance Jio, providing strong empirical substance to Reglitz claim is vital to understand. In September 2016, it entered the Indian market with data plans and handsets at nominal rates when India's mobile data consumption ranking was 155th. Within a span of 3 years, the company amassed nearly 370M subscribers~\cite{mint-reliance-2016} taking India to the first place. India's data usage per smartphone rose from 0.9GB in 2016 to 11GB per month in 2019~\cite{shona-ghosh-reliance-2016}. Jio built a 2.7 lakh kilometer optical fibre net for a cost of \$22B and offered users attractive Internet plans which were almost 30x cheaper than its competitors, alongwith no roaming charges and free calls. Jio was able to pull this off because once the expensive 4G network infrastructure was setup, running the network had relatively low operating costs. 

\textbf{Only 4G:} When most other telecommunication operators in India were focused on developing 3G in 2010, Jio made a rather audacious bet on 4G. Today, Jio operates a network that is exclusively 4G/LTE, giving it an advantage over its competitors. Jio simply needs to manage one network - the one for data; whereas others need to manage a network for voice as well as one for data. Compared to its competitors, Jio requires far less physical equipments. Additionally, the voice networks that others must manage use outdated technology, consume more energy, and take up more space. This significantly raises their cost, which Jio completely avoids\cite{jio-different}.

\textbf{Low-Cost Phone Calls:} Jio could practically offer phone calls for free due to their low network utilization. They do not incur any additional costs for managing a separate voice network. Additionally, unlike high data content like HD videos, phone conversations produce extremely few packets that need to be sent. Thus, they are able to offer unimaginably cheap plans\cite{jio-different}.

\textbf{Customer Retention:} Jio's strategy was always to draw customers in with affordable data, then hook them with entertainment apps like JioSaavn for music, JioCinema for movies, JioNews for news, etc. And finally, sell these users important services like internet shopping, education, healthcare, etc. to earn money off of them\cite{jio-different}.

Thanks to inexpensive Jio phones, even people with modest means, such as farmers, daily wage workers and low-income households, had access to the Internet. 

\subsection{Misinformation Harms} Providing a right would contextualise Internet usage within a legal framework and grant powers to penalise bad actors. The Internet no doubt has a plethora of harms as well as benefits. The question is whether provisioning a right would promote bad actors and the consequences thereafter? To answer these would require ensuring that Internet access is well protected and there are legal provisions in place to be able to provide a safe environment on the Internet.  With the development of a right, other auxiliary rights such as data protection would be necessitated. Protecting the right to Internet access would imply subsequent data protection and data privacy legislations. But from a technological point of view, the possibility of Internet harms should not deter us from providing Internet to all. Just because rash vehicle drivers can be harmful to the society, we do not discourage people from driving on the road.    

\subsection{Comparison with other Rights} The third question directly delves into some of the most influential arguments of Cerf - the need to have a high bar for something to be a human right. Cerf’s definitional distinction of a civil right and a human right leaves no room for the idea that human rights can change as social contexts change, and rather subtly suggests that Internet divides would not cause significant harm. Internet access is indeed a special right unlike other rights. It would be unfair to have similar expectations as other rights. 

Are our responsibilities for Internet access more significant than, say, making sure everyone has access to electricity? As Chike Aguh, CEO of EveryoneOn puts it, human rights are not inalienable\cite{websummit}. Prioritizing free speech over the right to education would not be a good idea. All rights coexist, are undeniably significant, and Internet access functions as a primary right that makes practically all other rights possible. But that doesn't mean we shouldn't give one right precedence over another. Such a conundrum might not exist in nations with 100\% electrification rates, such as India, the United States, Germany, and France, where electrification would not be a concern~\cite{electrification-us} and Internet connection would not be a potential issue. But in places like South Sudan, Chad or Liberia~\cite{electrification-rate}, electricity is imperative to achieve Internet access. So, while some regions would need prioritizing one right over another, it would be impossible to make an exclusive choice of electricity over Internet or vice versa.  

\section{Conclusion}
India's digital journey is crucial to steer this debate in the right direction. Studying the challenges of a digital democracy with large populations of both, Internet haves and have nots, provides us with important lessons weighing in favour of a human right. The children of Gadchiroli were denied access to education when their villages did not have Internet. The \textit{Faheema Shirin vs State of Kerala case of 2019} case providing a landmark judgment, recognized the importance of the right and cautioned us against the possibility of arbitrary revocation. The Anuradha Bhasin case underlined how treating Internet as just a means rather than an end can be easily exploited by substituting it with other less sophisticated alternatives. The cases also emphasized the need to look at the right from a more utilitarian perspective to be able to correctly prioritize it with other rights. 

The success story of Jio providing affordable internet to nearly 300 million people painted a rough picture of the duties that such a right would entail.

Cerf might be satisfied with India overly fulfilling its duty by provisioning a civil right, but Reglitz might argue for more effort to eliminate digital divide. Universal Internet access would not occur naturally without intentional promotion but Internet divides would. The human right to Internet access in the form of government legislation can put appropriate obligation on governments to rule out Internet divide, have a legal system tied around the same for safe and effective usage, and promote other derived rights.

\section{Acknowledgements}
We would like to specifically thank Kristin Williams, Alexis Newton and Deepak Dhole for providing thoughts and other students of the Social and Ethical Issues in Computing class at Emory University.

\bibliographystyle{ACM-Reference-Format}
\bibliography{sample-base}
\appendix

\end{document}